\documentclass[aps,prl,twocolumn,groupedaddress]{revtex4}

\usepackage{graphicx,bm}

\begin{document}

\preprint{Draft (\today)}

\title{Electric Control of Spin Helicity in a Magnetic Ferroelectric}

\author{Y. Yamasaki$^1$, H. Sagayama$^2$, T. Goto$^1$, M. Matsuura$^3$, K. Hirota$^3$, T. Arima$^{2,4}$ and Y. Tokura$^{1,4,5}$}
\affiliation{$^{1}$Department of Applied Physics, University of Tokyo, Tokyo 113-8656, Japan} 
\affiliation{$^{2}$Institute of Multidisciplinary Research for Advanced Materials, Tohoku University, Sendai 980-8577, Japan}
\affiliation{$^{3}$The Institute for Solid State Physics, University of Tokyo, Kashiwa 277-8581, Japan}
\affiliation{$^{4}$Spin Superstructure Project, ERATO, Japan Science and Technology Agency, 
Tsukuba 305-8562, Japan}
\affiliation{$^{5}$Correlated Electron Research Center (CERC), National Institute of Advanced 
Industrial Science and Technology (AIST), Tsukuba 305-8562, Japan}

\date{\today}

\begin{abstract}
Magnetic ferroelectrics or multiferroics, which are currently extensively explored, may provide a good arena to realize a novel magnetoelectric function.  Here we demonstrate the genuine electric control of the spiral magnetic structure in one of such magnetic ferroelectrics, TbMnO$_3$.  A spin-polarized neutron scattering experiment clearly shows that the spin helicity, clockwise or counter-clockwise, is controlled by the direction of spontaneous polarization and hence by the polarity of the small cooling electric field. 
\end{abstract}


\maketitle

Electric control of magnetic spins or their ordering structure has long been a big challenge in condensed matter physics. Furthermore, manipulating the magnetization by electric field may provide a low energy-consuming way in spin-electronics and a higher data density in information storages\cite{Fiebig,Tokura_Science}. There are a number of magneto-electric materials whose magnetization can be changed, though minutely, with an external electric field, yet only a very few are known whose magnetic structure itself can be controlled by an electric field\cite{Fiebig,Ascher,Lottemoser,Zhao}. The use of ferroelectricity is perhaps indispensable to enhance the electric field action on the magnetic spins.\cite{Tokura_Science}
 
 One of the robust mechanisms to produce the ferroelectricicty of magnetic origin has been recently proposed by Katsura, Nagaosa, and Balatsky (KNB) \cite{KNB}. The overlap of the electronic wave function between the two atomic sites ($i$ and $i + 1$) with mutually canted spins ($\bm{S}_i$ and $\bm{S}_{i+1}$) can generate electric polarization, $\bm{p}_i= A\bm{e}_{i,i+1}\times(\bm{S}_i\times \bm{S}_{i+1})$, where $\bm{e}_{i,i+1}$ denotes the unit vector connecting the two sites and $A$ is a constant determined by the spin exchange interaction and the spin-orbit interaction. (Note that the similar theoretical results have been obtained independently also in refs. \cite{Mostovoy,Sergienko}). In case the transverse-spiral (cycloidal) spin order is realized (Fig. 1(b)), the uniform spontaneous polarization is expected to emerge as the sum of the local polarization $\bm{p}_i$ in the direction perpendicular to the spiral propagation vector and the vector chirality, $\bm{C}\equiv \sum_i \bm{S}_i\times \bm{S}_{i+1}$ (ref. \cite{KNB}). This spin-driven ferroelectricity has recently been found in several transverse-spiral magnets such as TbMnO$_3$ (ref. \cite{Kimura}), Ni$_3$V$_2$O$_8$ (ref. \cite{Lawes}), MnWO$_4$ (ref. \cite{Taniguchi}), and also in a transverse cone-spiral magnet CoCr$_2$O$_4$ (ref. \cite{Yamasaki}). We report here the quantitative elucidation of such magnetically induced ferroelectricity in terms of the spin ellipticity as the order parameter and show the successful electric control between the clockwise (CW) and counter-clockwise (CCW) spin helixes. 

\begin{figure}[htbp]
\begin{center}
\includegraphics[width=0.45\textwidth,keepaspectratio=true]{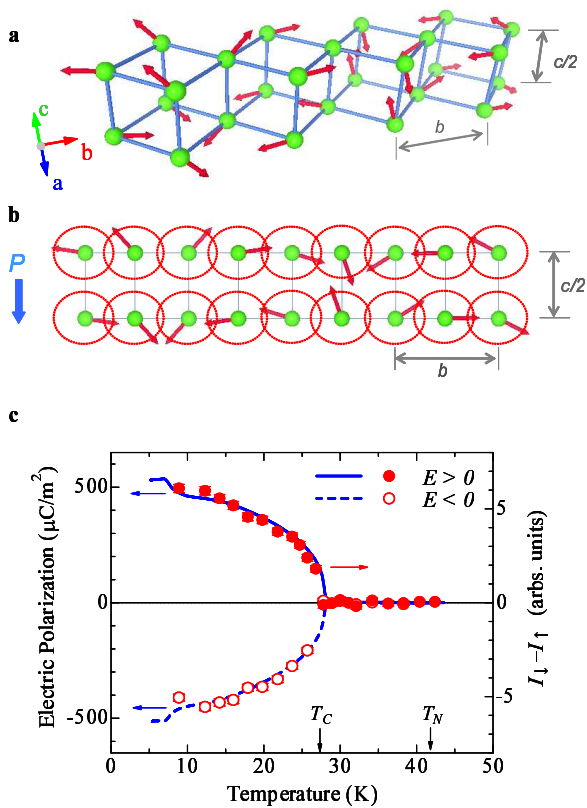}
\end{center}
\caption{(Color Online) (a) A schematic transverse-spiral (cycloidal) magnetic structure of TbMnO$_3$ ($a\times 4b \times c/2$ in $Pbnm$ orthorhombic cell) in the ferroelectric phase below $T_C = 27$ K. (b) A schematic magnetic structure projected onto $bc$-plane (including the both $a = 0$ and $a = 1/2$ planes). The electric polarization $\bm{P}$ emerges along the $c$-axis, that is the direction perpendicular to the spiral propagation vector and the spin vector chirality. (c) Temperature dependence of electric polarization along the $c$-axis and the difference of the intensities of magnetic satellites $(4,+q,1)$ when spins of the neutron beams is anti-parallel and parallel to scattering vector, $I_\downarrow-I_\uparrow$, with cooling electric field $E > 0$ and $E < 0$. }
\label{}
\end{figure}

A family of perovskite manganites, $R$MnO$_3$ with $R$ being Tb, Dy, and their solid solution, have recently been demonstrated to undergo a ferroelectric transition at the Curie temperature $T_C$ of $20-30$ K (see the example shown in Fig. 1(c)) \cite{Goto_PRL,RMnO3_Full}. Below $T_N \sim 40$ K, the compounds undergo a long-range spin ordering with the modulation vector $\bm{Q}_s=(0,\pm q,1)$ with $q = 1/2-1/4$ (in $Pbnm$ orthorhombic setting)\cite{Kimura,Quezel}. This has been ascribed to the spin frustration effect caused by the combination of GdFeO$_3$-type distortion and staggered orbital order (so-called C-type orbital order) \cite{Kimura,Kimura_Ishihara}. The ferroelectricity in this antiferromagnet is derived from the transverse-spiral or cycloidal spin ordering\cite{Kenzelmann,Arima} (Fig. 1(a) and (b)) in accord with the KNB model\cite{KNB}, being distinct from the collinear spin configuration at higher temperatures, $T_C < T < T_N$. In a single crystal of TbMnO$_3$ (orthorhombic with the space group $Pbnm$ at room temperature) investigated here, a sinusoidal incommensurate antiferromagnetic ordering of the Mn$^{3+}$ moments takes place at $T_N = 42$ K with a magnetic wave vector $q\sim 0.27 $\cite{Kimura,Quezel}. The ferroelectric polarization along the $c$-axis ($P_c$) emerges upon the magnetic phase transition from the collinear($\parallel b$) to the transverse spiral spin structure at $T_C = 27$ K (Fig. 1(c)).

The helicity of such a transverse-spiral magnetic structure can be ascertained by the polarized-neutron-diffraction intensity of magnetic satellite reflection\cite{Blume,Gukasov,Ishida,Shirane,Shiratori}. The spin helicity is determined by measuring intensities of two magnetic satellites with the polarized neutron spin $\bm{S}_n$ parallel and anti-parallel to the vector chirality $\bm{C}$. The first successful control of the spin helicity as demonstrated by a polarized neutron diffraction study was on a spinel crystal of ZnCr$_2$Se$_4$ which was cooled through the helical spin transition point in simultaneously applied magnetic and electric fields\cite{Shiratori}. This material exhibits a proper screw spin structure below 20 K, which can host no ferroelectric polarization of magnetic origin because of the spin helicity $\bm{C}$ being parallel to the modulation vector $\bm{Q}_m$. As a result, a magnetoelectric cooling procedure with relatively strong field magnitudes was necessary for the control of spin helicity. By contrast, the spontaneous polarization emerges in the present transverse-spiral magnet TbMnO$_3$ with $\bm{C}\perp \bm{Q}_m$ in which the spin helicity is expected to be controlled only with a weak cooling electric-field enabling the production of the single-domain ferroelectric state.

A single crystal of TbMnO$_3$ was grown by a floating-zone method in a flow of Ar gas. The crystal was cut into a thin plate with the widest face of (001) and a thickness of 2.5 mm. Aluminum electrodes were deposited onto the widest faces. Spin polarized neutron diffraction measurements were performed with a triple-axis spectrometer PONTA at JRR-3, Japan. The sample was mounted on a sapphire plate in a closed-cycle helium refrigerator and irradiated with a spin-polarized neutron beam. The collimation of the incident and scattered neutron beam was 40 min. A Heusler monochromator was utilized to obtain the spin-polarized neutron beam with a kinetic energy of 36.0 meV. The spin of the neutron beam could be flipped by a spin-flipper, so as to be parallel (when the flipper was off) or anti-parallel (on) to the scattering vector $\bm{Q}_s$ with a guide-field of about 1 mT applied by a Helmholtz coil, as depicted in Fig. 2(b). The spin flipping ratio of the incident neutron beam was 11.6(4). All the neutron diffraction measurements were performed without application of electric field after cooling the sample from 50 K in a poling field (160 kV/m). Peak profiles of magnetic satellite reflections $(4, +q, 1)$ and $(4, -q, 1)$ with $q \sim 0.27$ were obtained by rotating the sample around the vertical axis, which approximately corresponded to the $L$ scan in the reciprocal space (see Fig. 2(a)). For the measurement of electric polarization $\bm{P}$, pyroelectric current was measured with an electrometer (Keithley 6517A) in a warming run without the application of electric field after cooling the sample in a poling field (80 kV/m). 

Figure 2(c) shows the $L$-scan profiles of magnetic satellites at $\bm{Q}_s=(4,\pm q,1)$ for the ferroelectric state at 9 K which were measured after cooling in an electric field of $E = \pm 160$ kV/m as shown in Fig. 2(b). In the case of $P_c > 0$, the intensity of the satellite $(4,+q,1)$ is approximately 9 times as high with neutron spin ($\bm{S}_n$) anti-parallel ($I_\downarrow$) to the scattering vector $\bm{Q}_s$ as that with parallel $\bm{S}_n$ ($I_\uparrow$). As for the satellite $(4,-q,1)$, conversely, $I_\uparrow$ is much stronger than $I_\downarrow$. These behaviors are typical of a spiral magnet with a single helicity, where the spiral plane is almost perpendicular to the scattering vector. Here, it is to be noted that the spiral plane of TbMnO$_3$ is perpendicular to the $a$ axis. When the direction of the cooling electric field is reversed, the intensity $I_\uparrow$ of satellite $(4,+q,1)$ becomes much stronger than $I_\downarrow$, demonstrating that the opposite helicity domain becomes dominant. The reversal of the ratio of $I_\downarrow$ to $I_\uparrow$ induced by the reversal of the poling electric field suggests that the spin helicity can be successfully controlled by a poling electric field. 

\begin{figure}[htbp]
\begin{center}
\includegraphics[width=0.45\textwidth,keepaspectratio=true]{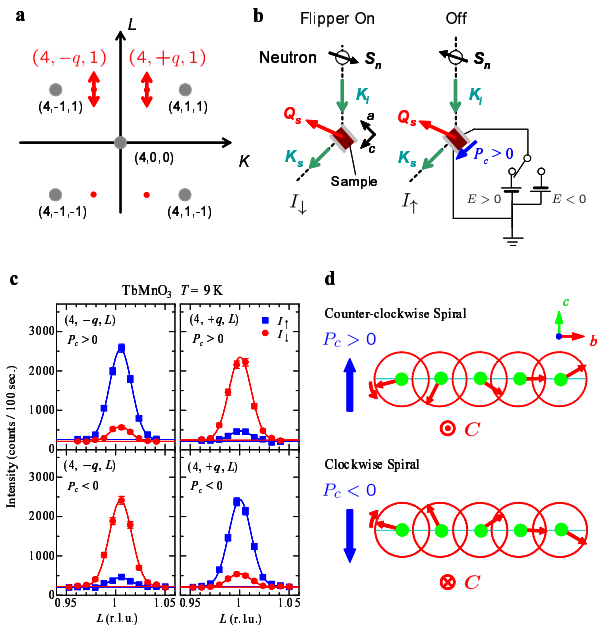}
\end{center}
\caption{(Color Online) (a) $L$-scans of magnetic satellites $(4,\pm q,1)$ with $q \sim 0.27$ in the present experiment as shown by red arrows in the reciprocal $(K, L)$-plane. (b) The schematic illustration of the apparatus. The spin of the neutron beam ($\bm{S}_n$) could be flipped by a spin-flipper so as to be parallel (when the flipper was off) or anti-parallel (on) to the scattering vector $\bm{Q}_s$. (c) Electric polarization dependence of intensity of magnetic satellites $(4,\pm q,1)$ at 9 K. $L$-scan profiles of magnetic satellites $(4,\pm q,1)$ with $P_c > 0$ and $P_c < 0$. (d) The relation between the spin rotatory direction (or helicity) and the direction of electric polarization in TbMnO$_3$. }
\label{}
\end{figure}

In the following, we consider the interrelation among the electric polarization, the spin helicity, and the cross section of the elliptical spiral magnetic structure. In the transverse-spiral magnetic phase below 27 K, the magnetic moment $M_i$ on the $i$-th Mn site is described as 
\begin{equation}
\bm{M}_i=\bm{m}_b \cos (2\pi \bm{Q}_m\cdot \bm{R}_i) + \bm{m}_c \sin (2\pi \bm{Q}_m\cdot \bm{R}_i).
\end{equation}
It was reported for TbMnO$_3$ that $\bm{m}_b \sim (0, 3.9, 0) \mu_B$ and $\bm{m}_c \sim (0, 0, 2.8) \mu_B$ at 15 K by a magnetic structure analysis\cite{Kenzelmann,Arima}. This magnetic modulation produces two satellite peaks around each fundamental reflection. The cross sections of the respective satellites $(4,\pm q,1)$ are calculated as 
\begin{equation}
\left(\frac{d\sigma}{d\Omega}\right)^\pm =I_o \left\{ m_b^2 + m_c^2 \mp 2m_bm_c (\hat{\bm{S}}_n \cdot \hat{\bm{C}} )\right\}.
\end{equation}
Here, $I_o$ is the proportional constant determined by the magnetic structural factor, $m_c =|\bm{m}_c|$, $m_b = |\bm{m}_b|$, and the hat symbols indicate the unit vectors. This cross section (2) is derived from the approximation that the scattering vector $\bm{Q}_s$ and the neutron spins $\bm{S}_n$ are parallel or anti-parallel to the vector chirality $\bm{C}$. In the case of CCW spiral magnetic structure (the $a$-component of vector chirality; $C_a>0$ ), the intensities of $(4,+q,1)$ are calculated by using Eq. (2); $I_\downarrow \sim I_o (m_c-m_b)$ and $I_\downarrow \sim I_o (m_c+m_b)$ for $\bm{S}_n$ being parallel and antiparallel to $\bm{Q}_s$, respectively. Provided that $m_c/m_b\sim 0.7$ and also taking the spin flipping ratio of the incident neutron beam into account, these intensities are in accordance with the observed relative satellite intensities at $(4,+q,1)$ for the electric polarization $P_c>0$. It is thus confirmed that the CCW spiral magnetic structure corresponds to the electric polarization $P_c>0$ and the CW to $P_c<0$ in TbMnO$_3$. 

We show in Fig. 3(b) the temperature dependence of the $Q$-integrated polarized-neutron-diffraction intensity of the magnetic satellite $(4,+q,1)$ in the $P_c > 0$ state. Note that the experimentally obtained intensities (plotted in Fig. 3(b)) were corrected with taking into account the imperfect spin polarization (about 92 \%) of the incident neutron beam prior to the calculation of ellipticity.  The magnetic satellite shows up below $T_N$, but there is no difference between $I_\downarrow$ and $I_\uparrow$ because of the sinusoidally modulated collinear spin structure. The difference between $I_\downarrow$ and $I_\uparrow$ emerges upon the ferroelectric phase transition at $T_C$ due to the CCW spiral magnetic structure. We show in Fig. 3(a) the ellipticity of the spiral magnetic structure of TbMnO$_3$, defined as the ratio of amplitude of the $c$-axis spin component to the $b$-axis spin component in antiferromagnetic modulation, $m_c/m_b$, is experimentally estimated to the first approximation, 
\begin{equation}
\frac{m_c}{m_b}=\frac{\sqrt{I_\downarrow}-\sqrt{I_\uparrow}}{\sqrt{I_\downarrow}+\sqrt{I_\uparrow}}.
\end{equation}
The small angle between the scattering vector and the vector chirality of the ellipsoids (10.6	 degree) is neglected in this estimate. The induced magnetic moments of Tb ions were not taken into account because they are reported to be aligned along the $a$-axis and insensitive to the neutron scattering with $\bm{Q}_s$ approximately parallel to the $a$-axis \cite{Kenzelmann}. The ellipticity remains zero in the paraelectric phase and develops below the ferroelectric phase transition temperature in accordance with the ellipsoidal spiral ordering. The average intensity  , corresponding to the intensity of unpolarized neutron diffraction, almost continuously changes even through the ferroelectric phase transition, while the ellipticity $m_c/m_b$ discontinuously changes from zero to a finite value. The ellipticity at 15 K derived from the present experiments, $m_c/m_b \sim 0.63(2)$, shows a good agreement with the value ($m_c/m_b \sim 0.72$) determined by the magnetic structure analysis \cite{Kenzelmann}, considering the assumption adopted in the calculation. 

The temperature dependence of the polarized neutron magnetic scattering intensities provides another clear evidence that the ferroelectric polarization is inseparably related with the cycloidal spin structure. The KNB model predicts that the ferroelectric polarization should be proportional to $m_bm_c$, 
\begin{equation}
P_c=\mp A \sin (2\pi qb) m_bm_c
\end{equation}
where $b$ is the lattice constant. We compare in Fig. 1(c) the temperature dependencies of the observed ferroelectric polarization and the differential intensity of satellite $(4,+q,1)$, $I_\downarrow - I_\uparrow = \pm 4 I_o m_bm_c$. An excellent agreement between the both temperature dependencies clearly confirms that the ferroelectric polarization in the transverse-spiral magnetic TbMnO$_3$ is proportional to $m_bm_c$, and hence can be explained with the KNB or related model. 
\begin{figure}[htbp]
\begin{center}
\includegraphics[width=0.45\textwidth,keepaspectratio=true]{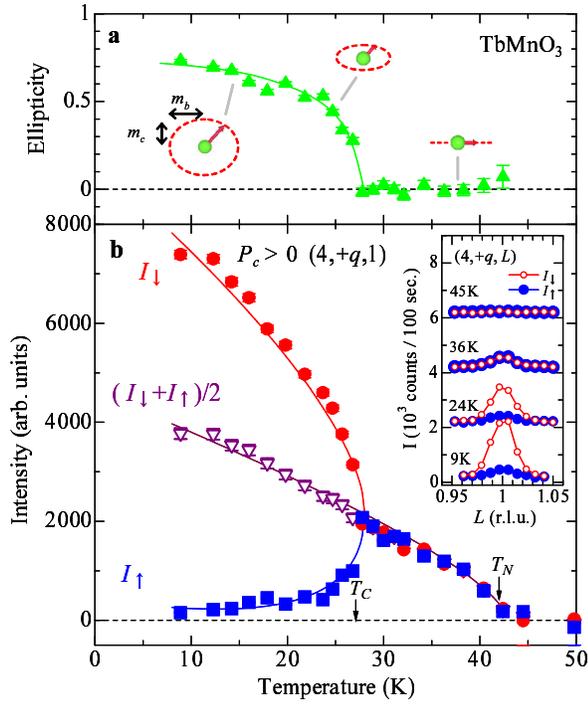}
\end{center}
\caption{(Color Online) (a) The temperature dependence of ellipticity, defined as $m_c/m_b$, as calculated by intensities of satellite reflections. (b) The temperature dependence of the $\bm{Q}$-integrated intensities at $\bm{Q}_s = (4, +q, 1)$, $I_\downarrow$ and $I_\uparrow$ for the neutron spin $\bm{S}_n$ parallel and antiparallel to $\bm{Q}_s$, and the average of these two intensities ($(I_\downarrow+I_\uparrow)/2$ ). These intensities are corrected by taking account of the spin flipping ratio of the incident neutron beam and the background intensities. The solid lines are the guide to the eyes. The inset shows the $L$-scan profiles at selected temperatures.  }
\label{}
\end{figure}

In summary, we have investigated the spin helicity of spiral magnetic structure in the magnetic ferroelectric TbMnO$_3$ by polarized neutron diffraction measurements. The intensities of satellite $(4,\pm q,1)$ with the neutron spin parallel ($I_\uparrow$) and antiparallel ($I_\downarrow$) to the scattering vector are observed to clearly differ in the ferroelectric phase. The reversal of poling electric field is observed to induce the reversal of magnitude relation between the intensities $I_\uparrow$ and $I_\downarrow$. This suggests that the spin helicity, clockwise or counter-clockwise, can be controlled by the poling electric field alone. The good agreement between the temperature dependencies of the electric polarization and the difference of satellite intensities ensures that the ferroelectric polarization in TbMnO$_3$ arises from the transverse-spiral (cycloidal) spin arrangement. Such a mechanism of generating electric polarization as based on the non-collinear magnetic ordering suggests further possibilities of electrically controllable magnetism. 

The neutron scattering experiments were carried out under the ISSP joint-research program. The authors are grateful to H. Katsura and N. Nagaosa for enlightening discussions. This work was supported in part by Grants-In-Aid for Scientific Research (Grant No.15104006, 16076207, 16076205 and 17340104) from the MEXT of Japan.


\begin{thebibliography}{20}
\bibitem{Fiebig} M. Fiebig, J. Phys. D: Appl. Phys. \textbf{38}, R123 (2005).
\bibitem{Tokura_Science} Y. Tokura, Science \textbf{312}, 1481 (2006).
\bibitem{Ascher} E. Ascher, H. Rieder, H. Schmid, and H. Stossel, J. Appl. Phys. \textbf{37}, 1404 (1966). 
\bibitem{Lottemoser} T. Lottermoser $et~al$, Nature (London) \textbf{430}, 541 (2004).
\bibitem{Zhao} T. Zhao, $et~al.$, Nat. Mat. \textbf{5}, 823 (2006).
\bibitem{KNB} H. Katsura, N. Nagaosa, and A. V. Balatsky, Phys. Rev. Lett. \textbf{95}, 057205 (2005).
\bibitem{Mostovoy} M. Mostovoy, Phys. Rev. Lett. \textbf{96}, 067601 (2006). 
\bibitem{Sergienko} I. A. Sergienko, E. Dagotto, Phys. Rev. B \textbf{73}, 094434 (2006). 
\bibitem{Kimura} T. Kimura $et~al.$, Nature \textbf{426}, 55 (2003).
\bibitem{Lawes} G. Lawes $et~al.$, Phys. Rev. Lett. \textbf{95}, 087205 (2005). 
\bibitem{Taniguchi} K. Taniguchi, N. Abe, T. Takenobu, Y. Iwasa, and T. Arima, Phys. Rev. Lett. \textbf{97}, 097203 (2006).
\bibitem{Yamasaki} Y. Yamasaki $et~al.$, Phys. Rev. Lett. \textbf{96}, 207204 (2006).
\bibitem{Goto_PRL} T. Goto, T. Kimura, G. Lawes, A. P. Ramirez, and Y. Tokura, Phys. Rev. Lett. \textbf{92}, 257201 (2004).
\bibitem{RMnO3_Full} T. Kimura, G. Lawes, T. Goto, Y. Tokura, and A. P. Ramirez, Phys. Rev. B. \textbf{71}, 224425 (2005).
\bibitem{Quezel} S. Quzel $et~al.$, Physica B \textbf{86-88},916 (1977).
\bibitem{Kimura_Ishihara} T. Kimura $et~al.$, Phys. Rev. B. \textbf{68}, 060403(R) (2003).
\bibitem{Kenzelmann} M. Kenzelmann $et~al.$, Phys. Rev. Lett. \textbf{95}, 087206 (2005). 
\bibitem{Arima} T. Arima $et~al.$, Phys. Rev. Lett. \textbf{96}, 097202 (2006).
\bibitem{Blume} M. Blume, Phys. Rev. \textbf{130}, 1670 (1963). 
\bibitem{Gukasov} A. Gukasov, Physica B \textbf{267-268}, 97-105 (1999).
\bibitem{Ishida} M. Ishida $et~al.$, J.Phys. Soc. Japan \textbf{54}, 2975 (1985).
\bibitem{Shirane} G. Shirane $et~al.$, Phys. Rev. B \textbf{28}, 6251 (1983).
\bibitem{Shiratori} K. Shiratori $et~al.$, J. Phys.Soc. Jpn. \textbf{48}, 1111 (1980).
\end{thebibliography}
\end{document}